\documentclass[epsf,twocolumn,showpacs,preprintnumbers]{revtex4}
\usepackage{graphics}
\usepackage{graphicx}
\usepackage{dcolumn} 
\usepackage{bm}
\usepackage{epsfig}
\begin{document}
\title{Magnetism Driven by Anion Vacancies in Superconducting 
 $\alpha$--FeSe$_{1-x}$}
\author{K.-W. Lee$^{1,2}$, V. Pardo,$^{1,3}$ and W. E. Pickett$^1$} 
\affiliation{ 
$^1$Department of Physics, University of California, Davis,
  CA 95616, USA \\
$^2$Department of Display and Semiconductor Physics, 
  Korea University, Jochiwon, Chungnam 339-700, Korea\\
$^3$Departamento de Fisica Aplicada, Facultad de Fisica, 
 Universidad de Santiago de Compostela, E-15782 Campus Sur s/n,
Santiago de Compostela, Spain
}
\date{\today}
\begin{abstract}
 To study the microscopic electronic and magnetic interactions 
 in the substoichiometric iron chalcogenide FeSe$_{1-x}$ 
 which is observed to superconduct at $x\approx \frac{1}{8}$ 
 up to $T_c$=27 K, we use first
 principles methods to study the Se vacancy in this nearly magnetic FeSe system.
 The vacancy forms a ferrimagnetic cluster of eight Fe atoms, which for the ordered $x$=$\frac{1}{8}$ alloy 
 leads to half metallic conduction.
 Similar magnetic clusters are obtained for FeTe$_{1-x}$ 
 and for BaFe$_2$As$_2$ with an As vacancy, although neither of these
 are half metallic. 
 Based on fixed spin density results, we suggest the low energy
 excitations in FeSe$_{1-x}$ are antiparamagnon-like with
 short correlation length.
\end{abstract}
\maketitle

\section{Introduction}
The discovery of superconductivity,\cite{hosono} 
now up to 55 K in the LaFeAsO class
of compounds, followed by T$_c$ of nearly 30 K in the BaFe$_2$As$_2$ class
with the same active conducting 
Fe$_2$As$_2$ layer has caused tremendous excitement.  A competition
between superconductivity and magnetism is very evident,
and the assortment of phenomena, concepts, and proposed models
is leading to a wide variety of suggestions that is reminiscent of the heyday of high
temperature superconducting cuprates.  
There is a serious need to identify and
address relatively straightforward questions, 
in addition to broader investigations
to compare and contrast all the ${\cal R}$FeAsO materials
(${\cal R}$=rare earth) to identify trends that might provide a
clue.  The system, FeSe, has the same band filling as the superconducting oxypnictides, 
but having only two atoms,
is structurally simpler and provides more direct questions.
Although many samples are two phase and are not always fully
characterized, this system is reported to be non-magnetic and
non-superconducting at stoichiometry, and is
magnetic and clearly superconducting \cite{taiwan,japan,uk,fang,yeh} 
at Se substoichiometry FeSe$_{1-x}$.

FeSe can be substoichiometric on either sublattice and contains
two major phases ($\alpha$ and $\beta$).
The PbO-type $\alpha$-FeSe$_{1-x}$ compound is the one of current
interest and has been studied extensively
for its spintronics-related magnetic properties by Shen 
and coworkers,\cite{feng,liu,wu1,wu2}
who concluded from observed hysteresis a ferromagnetic (FM) state 
in the nonstoichiometric phase, 
but a nonmagnetic state in the stoichiometric phase.
Very recently, Hsu {\it et al.} reported superconductivity with $T_c=8$ K,
at $x=0.12$ and 0.18.\cite{taiwan} 
Subsequently, $T_c$ has been raised rapidly up to 27 K 
at pressure $P=1.48$ GPa,\cite{japan,uk}
clearly putting FeSe$_{1-x}$ in the high T$_c$ category with 
iron pnictides having similar band filling.
Margadonna {\it et al.} confirmed $T_c \sim 14$ K at $x=0.08$
at ambient pressure.\cite{uk}
Fang {\it et al.} investigated isovalent 
Fe(Se$_{1-y}$Te$_y$)$_{0.82}$,\cite{fang}
finding $y$-dependent T$_c$ in the range of 8--14 K 
with maximum at $y \approx 0.6$, dipping to zero at $y=1$.
Notably, temperature dependent susceptibility measurements 
show an anomaly around 100 K, indicating a peculiar 
and not yet understood magnetic instability.\cite{taiwan,fang}
Hence, the competition between superconductivity and magnetism observed
in iron pnictides is clearly extended to these iron chalcogenides,
and the current picture seems to be that superconductivity arises
in a phase with strong magnetic character.
Here we focus on a crucial feature: Se vacancies are necessary for
producing the high temperature superconducting state,
so what is the character of this defect?

\begin{table*}[bt]
\caption{Optimized structure for a 2$\times$2 supercell 
with a Se vacancy (space group: $P4mm$, No. 99), {\it i.e.} Fe$_8$Se$_7$.
The order of Fe--Fe distances for the relaxed structure is
Fe1--Fe1, Fe2--Fe2, and Fe1--Fe2.
Fe1 means Fe atoms near a Se vacancy.
Note that this structure is optimized in ferromagnetic state.
}
\begin{center}
\begin{tabular}{ccccccccc}\hline\hline
     &    &\multicolumn{3}{c}{Unrelaxed} & &\multicolumn{3}{c}{Relaxed}\\
                                                   \cline{3-5}\cline{7-9}
     &    & $x$  & $y$  & $z$  &~& $x$ & $y$ & $z$   \\\hline
~~ Fe1 ~~ &~~ $4e$ ~~&~~ 0 ~~&~~ 0.25 ~~&~~ 0 ~~&~~ &~~ 0 ~~&~~ 0.2520 ~~&~~ 0.2375~~ \\
 Fe2 & $4f$ & 0.5 & 0.25 & 0 & & 0.5 & 0.2637 & 0.2350 \\
 Se1 & $2c$ & 0.5 & 0 & 0.7628 & & 0.5 & 0 & 0.9950 \\
 Se2 & $4d$ & 0.25 & 0.25 & 0.2372 & & 0.2481 & 0.2481 & 0.4795 \\
 Se3 & $1a$ & 0 & 0 & 0.7628 & & 0 & 0 & 0.0020 \\ \hline
 \multicolumn{2}{c}{bond}& \multicolumn{3}{c}{Fe--Se: 2.28}& &
 \multicolumn{3}{c}{Fe1(Fe2)--Se: 2.33(2.29)} \\
 \multicolumn{2}{c}{length (\AA)} &
 \multicolumn{3}{c}{Fe--Fe: 2.66} & &
 \multicolumn{3}{c}{Fe--Fe: 2.52, 2.68, 2.73} \\\hline\hline
\end{tabular}
\end{center}
\label{table1}
\end{table*}

\section{Results}
\begin{figure*}[tbp]
\flushleft
{\resizebox{7.7cm}{5.1cm}{\includegraphics{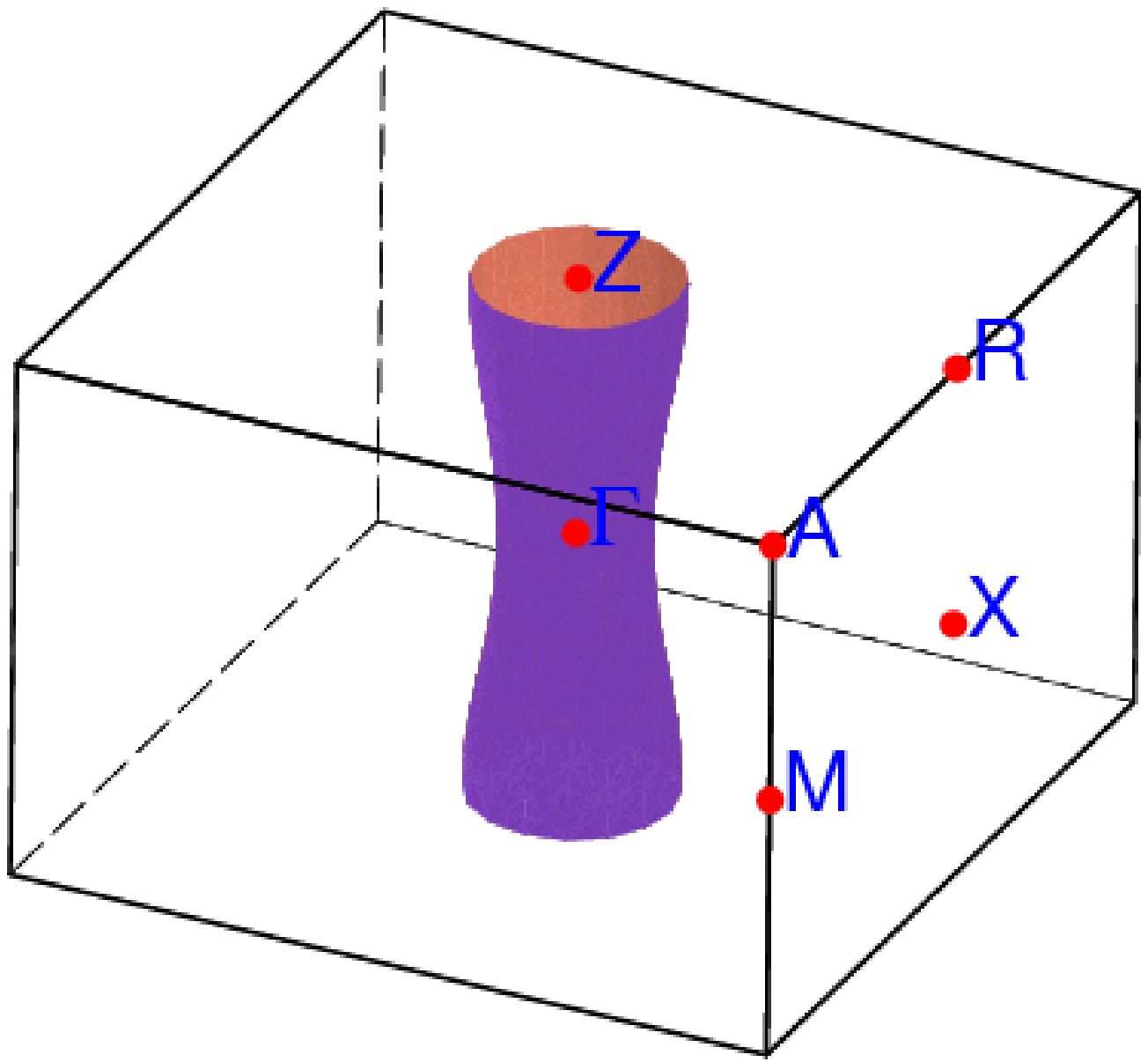}}}
\flushright
\vskip -55.0mm
{\resizebox{7.7cm}{5.1cm}{\includegraphics{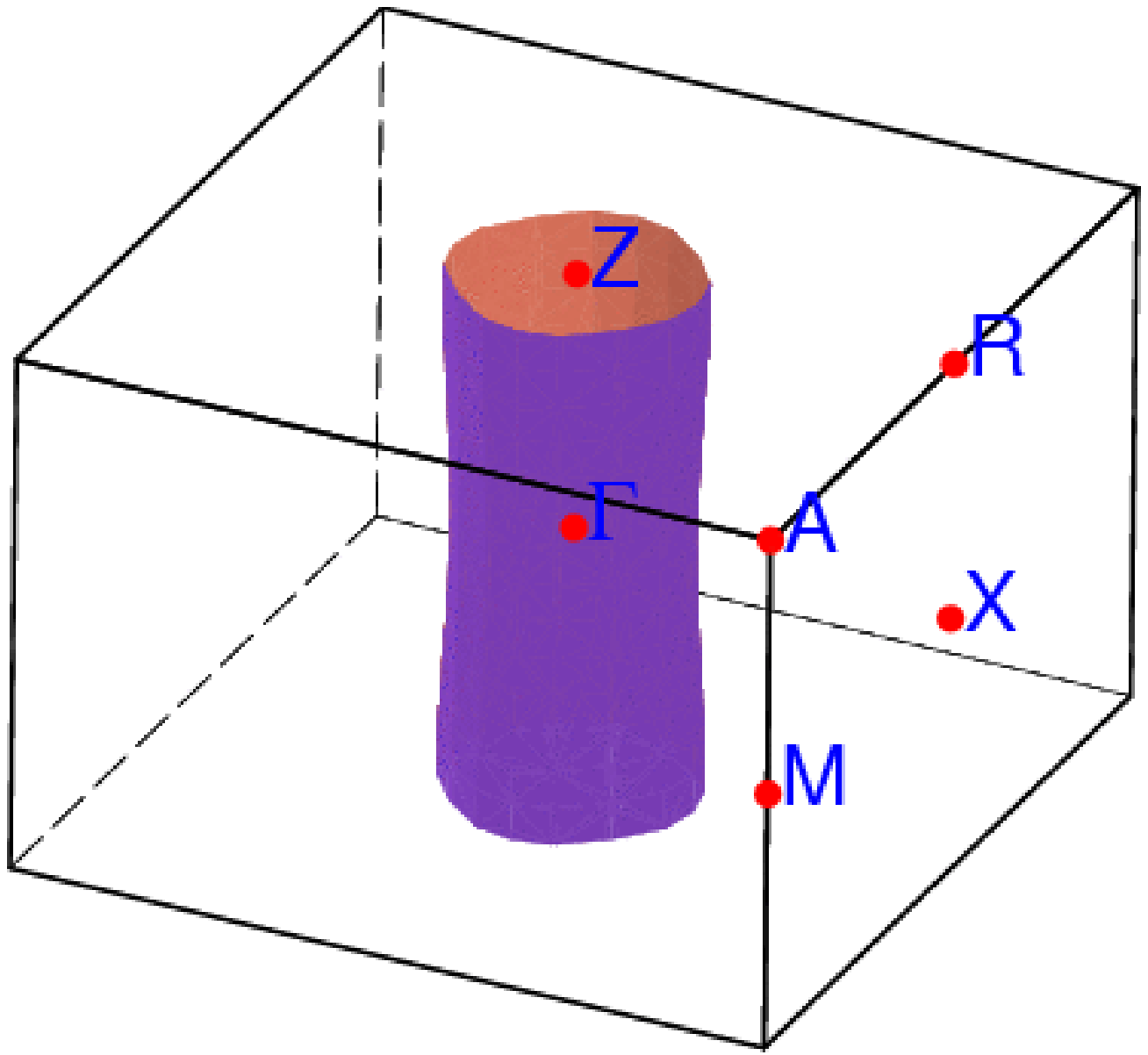}}}
\vskip 4mm
\flushleft
{\resizebox{7.7cm}{5.1cm}{\includegraphics{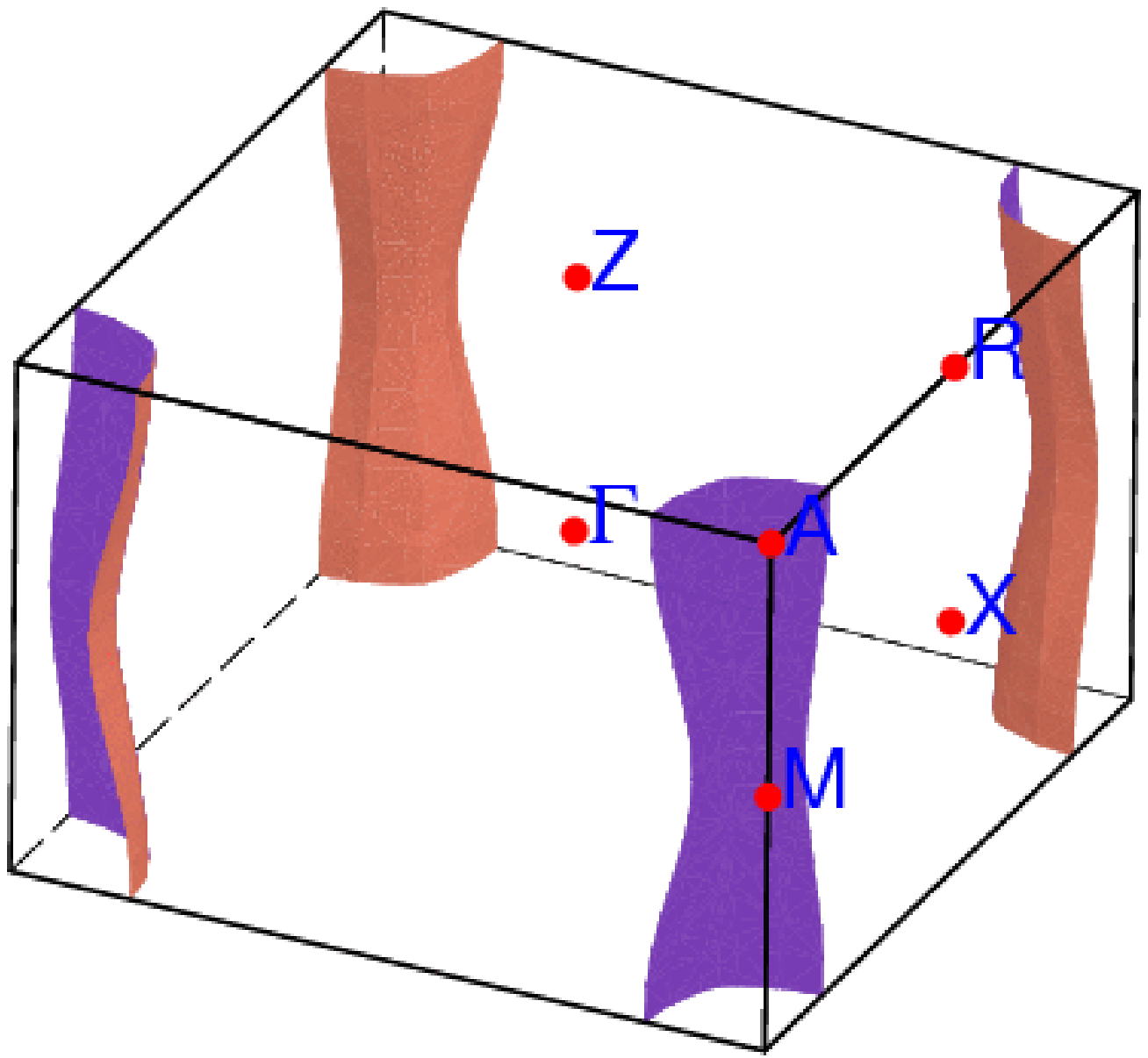}}}
\flushright
\vskip -55mm
{\resizebox{7.7cm}{5.1cm}{\includegraphics{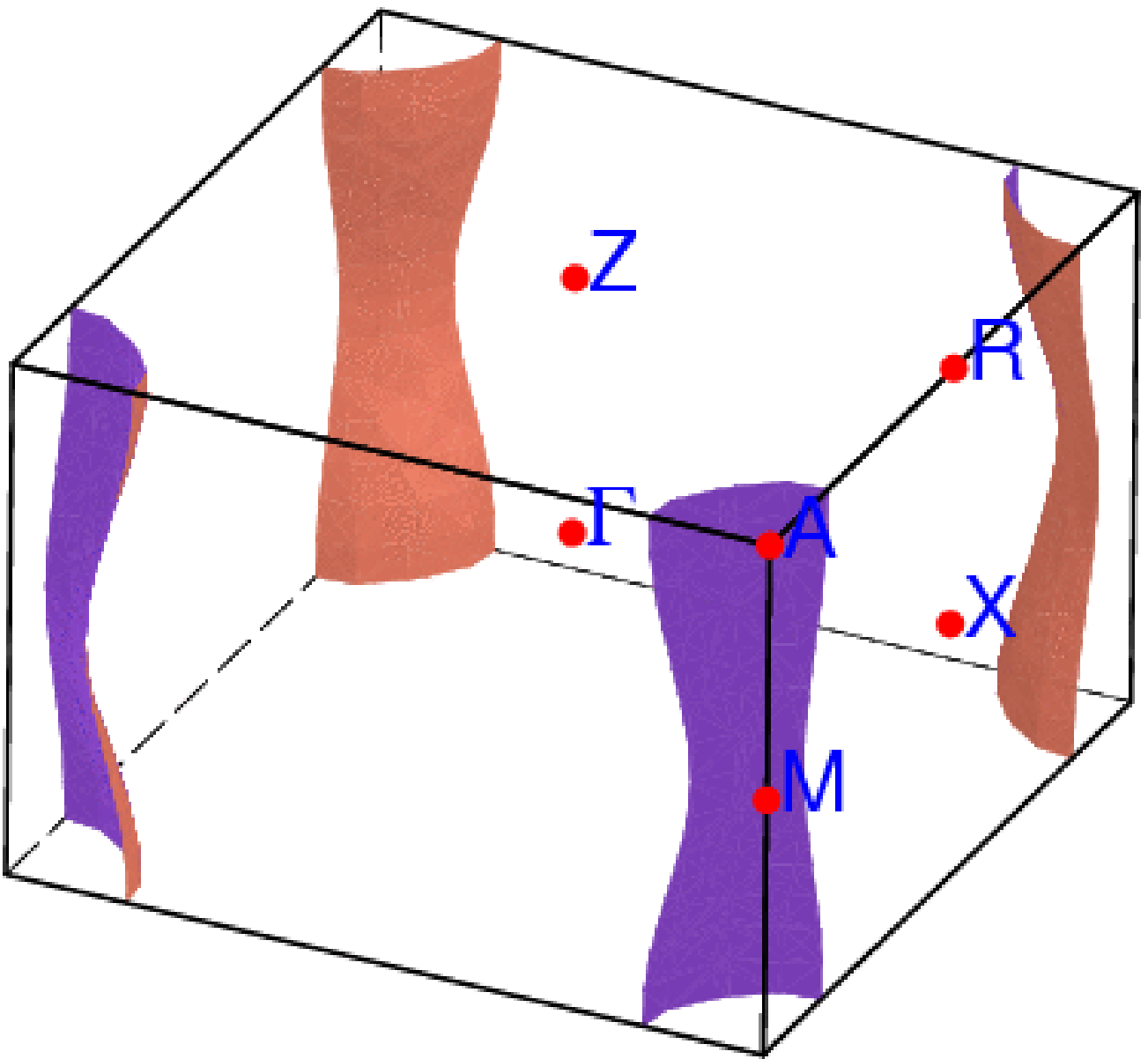}}}
\vskip 4mm
\caption{(Color online) Fermi surfaces (FSs) for nonmagnetic $x=$0
phase.
The FSs, consistent with those shown already
by Subedi {\it et al.},\cite{subedi}
consist of two $\Gamma$-centered hole cylinders,
which contain 0.07 and 0.04 holes per a Fe respectively,
and two compensating $M$-centered electron cylinders
with more dispersion along the $k_z$ direction.
}
\label{FS}
\end{figure*}

\begin{figure}[tbp]
{\resizebox{8cm}{6cm}{\includegraphics{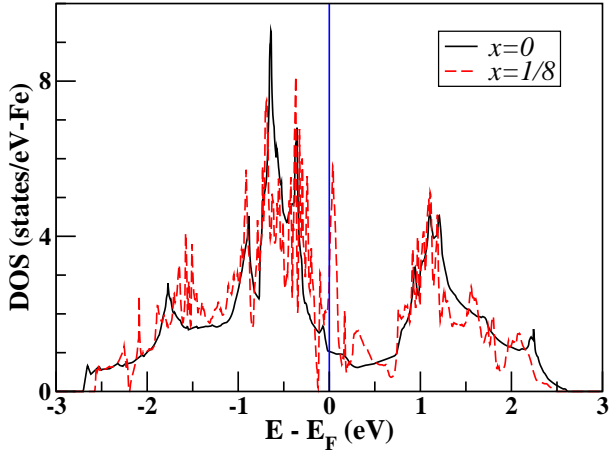}}}
\caption{(Color online) Total densities of states (DOSs)
 per Fe (both spins), for nonmagnetic $x$=0 and nonmagnetic
 $x=\frac{1}{8}$ in the optimized structures,
 in the regime of Fe $3d$ states.
 Note that the Fermi level $E_F$ of $x=\frac{1}{8}$ lies on steep
 side of a sharp peak, promoting a stable magnetic state.
 $N(E_F)$ of $x=\frac{1}{8}$ at $E_F$ is 3.07 states/eV per one Fe,
 which is about 3 times larger than that of $x=0$.
}
\label{pmdos}
\end{figure}

\section{Structure and calculation}
In the tetragonal phase $\alpha$-FeSe with PbO (B10) structure
(space group: $P4/nmm$, No. 129), Fe and Se atoms lie at $2a$ sites
(0,0,0) and at $2c$ sites (0,$\frac{1}{2}$,$z$), respectively.
The Fe$_2$Se$_2$ layers have the same structure as in 
LaFeAsO and BaFe$_2$As$_2$.
The experimental lattice parameters $a$=3.7693 \AA~ and 
$c$=5.4847 \AA~, which are reported recently by Hsu {\it{et al.}}
at $x$=0.12,\cite{taiwan} are used in our calculations.
The internal parameter $z=0.2372$ is optimized by energy minimization
within the local density approximation (LDA). 

Since the superconductivity has been observed around $x=0.12$ at ambient
pressure, we have used a 2$\times$2 supercell containing 8 formula units.
In this supercell, a Se vacancy represents the $x=0.125$ phase, 
well within the superconducting regime.
In this phase, this supercell contains two types of Fe atoms, one being 
adjacent to the Se vacancy (Fe1) and the other farther away (Fe2).
As shown in our optimized structure given in Table I,
the main effect of the Se vacancy on the structure is to shift Fe1 atoms 
toward the vacant site, whereas Fe2 atoms are affected little.
The changes in interatomic distance are $-0.14$ \AA~ for Fe1--Fe1,
$+0.07$ \AA~ for Fe1--Fe2, and $+0.05$ \AA~ for Fe1--Se.

Two all-electron full-potential codes, FPLO-7\cite{fplo1,fplo2} 
and WIEN2k\cite{wien2k} based on the augmented plane wave$+$local orbitals
(APW$+$lo) method,\cite{sjo} have been used in these calculations, with
consistent results.
The Brillouin zone was sampled with regular dense mesh containing up to
720 irreducible $k$ points.
Using WIEN2k with the Perdew-Wang LDA exchange-correlation functional,\cite{lda}
the atomic positions in the 2$\times$2 supercell 
with one Se vacancy were optimized until forces were smaller 
than 2 mRy/a.u..
For WIEN2k, local orbitals were added to gain flexibility in dealing 
with semicore states, Fe $3p$ and Se $3d4s$. 
The basis size was determined by R$_{mt}$K$_{max}$= 6. 
Atomic radii used were 2.21 a.u. for Fe and 1.96 a.u. for Se.

\section{Results}

\subsection{Stoichiometric phase.}
All attempts to obtain either FM
or antiferromagnetic (AFM) states led only 
to nonmagnetic (NM) solution once the Se position is optimized.
Even in the fixed spin moment calculations, no (meta)stable
FM state is obtained. The nonmagnetic ground state at $x=0$
is consistent with experimental observations.\cite{wu2}
Subedi {\it et al.}\cite{subedi} reported an AFM ground state
with a very small stabilization energy; this result is not
in serious conflict since energies and moments in these materials 
are known to be very sensitive to structural and computational 
details.\cite{zhiping} We have found such an AF ordering is stable 
using the generalized gradient approximation, but not with our LDA approach 
(both with {\sc WIEN2k} and {\sc FPLO}).
All these results taken together indicate FeSe is very near a magnetic
critical point.

As seen in other superconductors containing FeAs layers,
a transition from tetragonal to a low temperature orthorhombic structure
has been observed around $T=70$ K.\cite{taiwan,uk}
As expected from the tiny changes in crystal structures,
the calculated change in electronic structures is slight. 
We will address only the tetragonal phase here.

The main difference in the band structure of FeSe \cite{subedi}
with respect to iron pnictide compounds with similar structure occurs 
along the $\Gamma-Z$ line ($k_z$ direction). FeSe has 
a similar band structure to BaFe$_2$As$_2$\cite{singh_bafeas}, 
with a flat band (a $d$ orbital lying in the $xy$-plane 
with a large Fe--Fe hopping integral) (at $-20$ meV) just below E$_F$, 
whereas this band lies above E$_F$ in LaOFeAs\cite{zhiping}, 
FeTe\cite{subedi} and LiFeAs\cite{singh_bafeas}.
As might be expected, the Fermi surfaces shown in Fig. \ref{FS} 
are less two-dimensional in FeSe than in iron pnictide compounds, 
which have another layer of atoms between FePn layers.

Figure \ref{pmdos} shows the total density of states for $x=0$ 
compared with that for $x=0.125$.  
The density of states $N(E_F)$ for $x=0$ is small,
25\% less than the value for LaFeAsO and providing no tendency for
a FM instability. However, there is a van Hove singularity 
at $-50$ meV, which is absent in the Fe--As superconductors.
A difference compared to Fe--As compounds is the
hybridization gap at $-3$ eV that suggests strong Fe--Se hybridization, 
consistent with about 5\% smaller Fe--Se distance in this compound 
than Fe--As distance in either LaFeAsO or BaFe$_2$As$_2$.
The Se $p$ states also lie somewhat lower than the As $p$ states.

\begin{figure}[tbp]
{\resizebox{8cm}{6cm}{\includegraphics{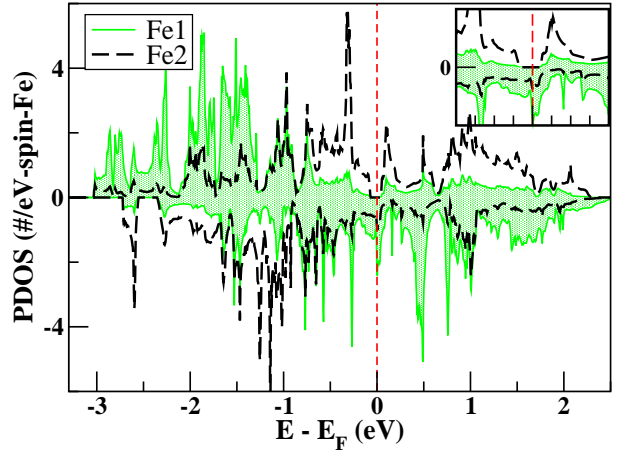}}}
\caption{(Color online) Fe atom-projected densities of state (PDOSs)
 of magnetic $x=\frac{1}{8}$ in the regime of Fe $3d$ states.
 The vertical line indicates E$_F$, set to zero,
 which passes through a peak in the minority and
 nearly bisects the gap of $\sim$0.1 eV in the majority.
 {\it Inset:} Blowup PDOS in the range of $-0.4$ to 0.4 eV, clearly showing
 a gap in the majority channel.
}
\label{fmdos}
\end{figure}

\begin{figure}[tbp]
{\resizebox{8cm}{6cm}{\includegraphics{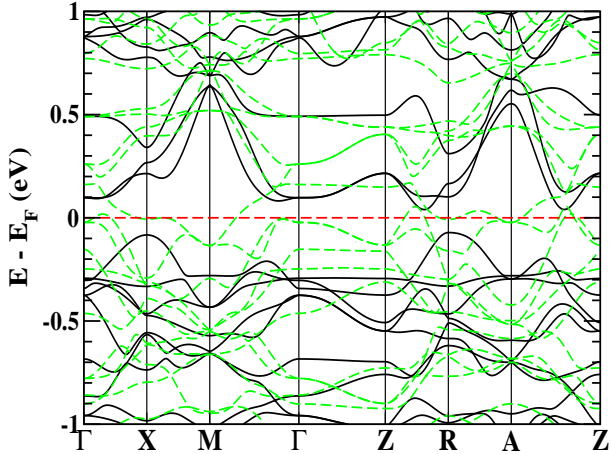}}}
\caption{(Color online) Blowup band structure of magnetic 
 $x=\frac{1}{8}$, showing half-metallicity with total spin 
 moment of 0.5 $\mu_B$ per Fe, near $E_F$ which is set to zero.
}
\label{fmband}
\end{figure}

\subsection{Effect of a Se vacancy}
This Se vacancy cannot be treated well 
by the virtual crystal approximation, which we have confirmed
by calculations, thus necessitating a supercell treatment of 
the actual vacancy. The virtual crystal approximation replaces both Se atoms and the vacancy by a peculiar average entity which artificially restores periodicity. Such a model cannot bear any relation to the defected system we study.
For $x=0.125$, we find that a strong magnetic state centered
on the vacancy is stabilized (see below).
This relaxed magnetic defect, with structure given in Table I, 
gains 32 meV/Fe by structural relaxation.
Disregarding magnetism, both structures (relaxed and unrelaxed) are nearly degenerate.
This difference reflects important magnetostructural coupling,
as already found for LaFeAsO,\cite{zhiping} and
dependence of optimized structure on magnetic states
is observed commonly in the FeAs--based superconductors,\cite{mazin,yildirim} 
which have large calculated spin moment.
In this relaxed structure, the magnetization energy (energy difference
between NM and FM) is 133 meV/Fe, one-third larger than 
in the unrelaxed structure and reflecting a very strongly magnetic cluster.

Now we will address the unusual properties of the magnetic state.
The most interesting point is that the two types of Fe ions are aligned 
antiparallel: Fe1 with 2.14 $\mu_B$ and Fe2 with $-1.10$ $\mu_B$
in the relaxed structure, which reflects a local AFM coupling 
(as opposed to a delocalized spin density wave mechanism). 
This distribution of Fe $d$ states of each spin can be observed 
in the atom-projected densities of states given in Fig. \ref{fmdos}.
The net moment is 0.5 $\mu_B$/Fe, and in addition 
this ordered system is half-metallic (1 $\mu_B$/Fe pair), as shown clearly in
the band structure given in Fig. \ref{fmband}.
In the unrelaxed structure each Fe ion has smaller moment 
in magnitude by 0.25 $\mu_B$, although the total moment remains unchanged.
The Se vacancy leads to creation of antialigned spin moments rather than 
any identifiable charge difference between Fe ions. 

\begin{figure}[tbp]
{\resizebox{8cm}{6cm}{\includegraphics{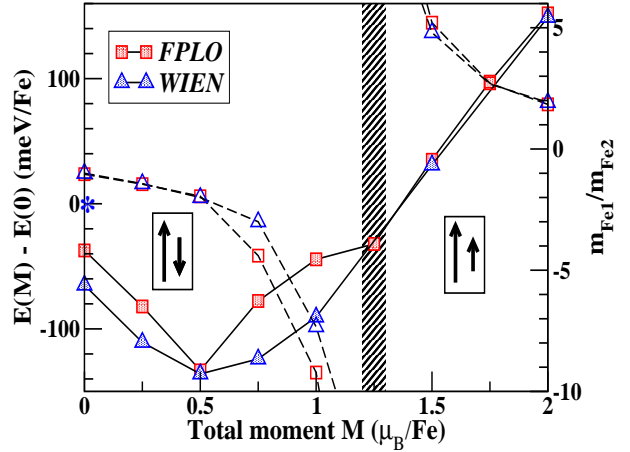}}}
\caption{(Color online) Total energy differences 
(solid lines, scale on the left axis) 
and ratio of local moments on Fe ions (dashed lines, scale on the right axis)
versus fixed spin moment, using FPLO and WIEN2k.
For $M\approx$0.75--1 $\mu_B$, two distinct states can be found.
The dashed area is a boundary separating regimes of 
antiparallel and parallel spin moments (at the boundary, 
Fe2 has nearly zero moment), described by arrows.
The symbol $\ast$ at $M$=0 denotes nonmagnetic state, which has
zero energy $E(0)$ in this plot.
The $M$=0 state at $-65$ meV has compensating moments on Fe1 and Fe2
of magnitude $\sim$1.9 $\mu_B$.
(For details, see text.)
}
\label{fsm}
\end{figure}

\begin{table*}[bt]
\caption{Optimized structure for a 2$\times$2 supercell
with an As vacancy (space group: $P4/mmm$, No. 123),
{\it i.e.} Ba$_4$Fe$_8$As$_7$, in ferromagnetic state.
The lattice parameters\cite{struct122}
$a$=3.9625 \AA~ and $c$=12.0168 \AA~ were used for the optimization.
In addition to atoms given below, Ba atoms lie at $1a$ site (0,0,0),
$1c$ site ($\frac{1}{2}$,$\frac{1}{2}$,0), and $2f$ sites (0,$\frac{1}{2}$,0).
The order of Fe--Fe distances for the relaxed structure is
Fe1--Fe1, Fe2--Fe2, and Fe1--Fe2.
Fe1 means Fe atoms near an As vacancy.
}
\begin{center}
\begin{tabular}{ccccccccc}\hline\hline
     &    &\multicolumn{3}{c}{Unrelaxed} & &\multicolumn{3}{c}{Relaxed}\\
                                                   \cline{3-5}\cline{7-9}
     &    & $x$  & $y$  & $z$  &~& $x$ & $y$ & $z$   \\\hline
~~ Fe1 ~~&~~ $8s$ ~~&~~ 0 ~~&~~ 0.25 ~~&~~ 0.25 ~~&~~ &~~ 0 ~~&~~ 0.2277 ~~&~~ 0.2561 ~~\\
 Fe2 & $8t$ & 0.25 & 0.5 & 0.25 & & 0.2506 & 0.5 & 0.2522 \\
 As1 & $4i$ & 0 & 0.5 & 0.6455 & & 0 & 0.5 & 0.6556 \\
 As2 & $2h$ & 0.5 & 0.5 & 0.6455  & & 0.5 & 0.5 & 0.6536 \\
 As3 & $8r$ & 0.25 & 0.25 & 0.1455 & & 0.2526 & 0.2526 & 0.1569 \\
 Ba & $4k$ & 0.25 & 0.25 & 0.5 & & 0.2548 & 0.2548 & 0.5 \\\hline
 \multicolumn{2}{c}{bond}& \multicolumn{3}{c}{Fe--As: } 2.40 & &
 \multicolumn{3}{c}{Fe1(Fe2)--As: 2.42 (2.32)} \\
 \multicolumn{2}{c}{length (\AA)} &
 \multicolumn{3}{c}{Fe--Fe: } 2.80 & &
 \multicolumn{3}{c}{Fe--Fe: } 2.55-2.80-2.93)\\\hline\hline
\end{tabular}
\end{center}
\label{table2}
\end{table*}

\subsection{Fixed spin moment studies}
One may ask: how stable is this antialigned local spin state? 
Fixed spin moment calculations\cite{fsm}
in the Fe$_8$Se$_7$ compound are used to investigate this question.
Results can be seen in Fig. \ref{fsm}, which shows the expected 
energy minimum around the half-metallic solution 
with 0.5 $\mu_B$/Fe (4 $\mu_B$ per magnetic cluster). 
Note that the E($M$) curve is not smooth at a point of half
metallicity\cite{scoo} where a substantial range of ``applied field"
leads to the same unchanging moment.
For total moment $M$=0, two solutions are found:
the antialigned spin state with net zero moment and 
the simple nonmagnetic state.
The antialigned state has lower energy by 65 (37 in FPLO) meV/Fe than
the nonmagnetic state, but higher energy by 71 meV/Fe than the half
metallic state.
However, the nonmagnetic solution is more stable than any solution 
with parallel aligned spins. 
These results indicate that parallel spins are strongly antagonistic
for this magnetic cluster; the local coupling is AFM.

For antialigned spin states, the difference in total energy 
between the two codes used here is associated with 
different local moments on Fe ions. These differences between two all-electron, full potential codes which usually give equivalent results reaffirm the strong sensitivity of the system to small effects.
At $M$=0 and 0.25 $\mu_B$, the moment of Fe1 obtained in WIEN2k
is about 15\% larger than in FPLO, though the ratio of Fe local moments
is nearly identical.
These differences probably reflect the sensitivity of FeSe$_{1-x}$--the
softness of its magnetism--to small computational details rather 
than representing distinct magnetic states.
Additionally, changing the total moment for the antialigned spin states,
the Fe1 moment is insensitive, with only a maximum change of 10\%, 
whereas the Fe2 moment varies rapidly.
From such behavior we can conclude that low energy excitations involve 
essentially little change in the Fe1 magnetic moment.

\subsection{Analogies in BaFe$_2$As$_2$ and FeTe}
To check the robustness of this vacancy induced magnetic cluster,
we carried out analogous calculations for FeTe$_{1-x}$
and BaFe$_2$As$_2$.
A similar structural relaxation was performed for 
FeTe$_{0.875}$ as in FeSe,\cite{fete} and
an As vacancy in BaFe$_2$As$_2$, {\it i.e.} Ba$_4$Fe$_8$As$_7$
given in Table \ref{table2}.
For BaFe$_2$As$_{1.75}$ and FeTe$_{0.875}$, 
the relaxation and magnetic cluster are similar
although the final states are not half metallic, 
with total moment of 0.7 $\mu_B$/Fe and 0.42 $\mu_B$/Fe, respectively. 
These magnetic states are favored energetically over
the nonmagnetic state by about 160 meV/Fe for BaFe$_2$As$_{1.75}$
and 175 meV/Fe for FeTe$_{0.875}$.

\section{Discussion and summary}
Now we consider the broader context.
Magnetism in superconducting samples, and its possible connection 
to superconductivity, is one of the primary
issues in iron-pnictide superconductivity, and our calculations
establish that Se (or Te, or As) vacancies promote 
strong magnetic clusters
surrounding the vacancy.  Superconductivity occurs only in
substoichiometric samples, and we obtain strong magnetic behavior
only around Se vacancies.
Our fixed spin moment results indicate the low energy excitations
will involve fluctuations in the magnitude of the next neighbor 
Fe spin (relative to the vacancy), while the near neighbor spin
remains rigid and antialigned.
The character of this excitation is antiparamagon-like but 
with short correlation length, a scenario 
that also seems relevant for the iron pnictide superconductors.

In the superconducting FeSe$_{0.88}$ materials there is a
magnetic transition characterized by a sharp upturn in the
susceptibility (apparently also with a structural aspect) near
105 K, followed by another transition at 75 K where the susceptibility 
abruptly returns to its higher temperature
value.\cite{taiwan,fang}  These anomalies have not
been discussed much yet, but the strong magnetic character,
and the difference in field-cooled and zero-field-cooled
susceptibility at lower temperature may be reflecting complex
cluster-glass behavior arising from the immobile but interacting magnetic
defects that we have studied.  The appearance of superconductivity
in a disordered magnetic system such as this provides strong
justification for further study of the physics of the FeSe$_{1-x}$
system.

\section{Acknowledgments}
We acknowledge S.-G. Lee for useful discussion about samples
and S. Leb$\grave{e}$gue for useful communications.
This research was supported by Korea University Grant No. K0718021 (K.W.L.),
Xunta de Galicia Human Resources Program (V.P.), 
and DOE Grant DE-FG03-01ER45876 (W.E.P.), 
and the work benefited from interactions within DOE's Computational
Materials Science Network.


\begin{thebibliography}{10}
\bibitem{fplo2} FPLO-5 has been used partially to consider
 the virtual crystal approximation (VCA) that has not been implemented
 in FPLO-7, yet.
 Valence orbitals included Fe $3s3p4s4p3d$ and Se $3p4s4p3d$ in FPLO-5.
\bibitem{fete} The experimental lattice parameters\cite{yeh}
$a$=3.819 \AA~ and $c$=6.250 \AA~ were used.
\end{thebibliography}
\end{document}